\documentclass{vgtc}                          

\graphicspath{{figures/}{pictures/}{images/}{./}} 

\usepackage{times}                     

\usepackage{tabu}                      
\usepackage{booktabs}                  
\usepackage{lipsum}                    
\usepackage{mwe}                       

\usepackage{mathptmx}                  

\onlineid{0}

\vgtccategory{Research}

\vgtcinsertpkg

\title{Focus360: Guiding User Attention in Immersive Videos for VR}

\author{
    Paulo Vitor S. Silva\thanks{e-mail: paulosantana@discente.ufg.br}, %
    Lucas L. Neves\thanks{e-mail: lucas.neves@discente.ufg.br},
    Rafael A. Goiás\thanks{e-mail: rafael\_goias@discente.ufg.br}, %
    Diogo F.C. Silva\thanks{e-mail: diiogofernands@gmail.com}, %
    Rafael T. Sousa\thanks{e-mail: rafaelsousa@ufmt.br}, %
    Arlindo R. Galvão Filho\thanks{e-mail: arlindogalvao@ufg.br} \\
    \parbox{\textwidth}{\scriptsize \centering Advanced Knowledge Center for Immersive Technologies, AKCIT \\ Federal University of Goiás, UFG}
}

\abstract{
    This demo introduces Focus360, a system designed to enhance user engagement in 360º VR videos by guiding attention to key elements within the scene. Using natural language descriptions, the system identifies important elements and applies a combination of visual effects to guide attention seamlessly. At the demonstration venue, participants can experience a 360º Safari Tour, showcasing the system's ability to improve user focus while maintaining an immersive experience.
} 


\begin{document}


\firstsection{Introduction}
\maketitle

Guiding users' attention involves identifying relevant information and directing focus toward particular elements within a visual scene \cite{silva2024attention}. With their immersive experience and 360-degree perspective, VR videos have become increasingly popular in various domains \cite{macquarrie2017cinematic}, including entertainment and tourism \cite{van2017virtual, dutta2024examining}. Virtual environment users have the freedom to explore the entire scene, potentially resulting in experiences that diverge from the intended road map \cite{mti6070054}. Nonetheless, the development of an effective narrative language for such media remains an area of active research \cite{maranes2020exploring}.

Several studies have investigated strategies for optimizing the guidance of users' attention through the use of visual features \cite{speicher2019exploring, schmitz2020directing}. The work of Silva et al. \cite{silva2024attention} proposes an approach that, through a video script and deep learning, automatically recognizes important elements in a scene and then guides the user’s attention to them through a vignette effect. 
However, the application of the vignette effect, while generally effective in guiding user attention, proves insufficient in certain scenarios. Specifically, when users are looking in the opposite direction of the element of interest, the effect fails to fulfill its purpose. Under such conditions, users perceive the majority of the screen as obscured or entirely black, making it challenging for them to discern where they need to look.

In this context, we propose a novel approach\footnote{Online Demonstration: \url{https://youtu.be/PSNaqcoIgDY}} that leverages a 360º video and a corresponding natural language description of its roadmap to automatically identify key elements and significant moments requiring user attention. By employing a combination of different visual effects, the method effectively guides user focus toward these regions of interest, enhancing the overall viewing experience and comprehension.

\section{System Architecture}
\begin{figure}[h]
  \centering
  \includegraphics[width=.8\columnwidth]{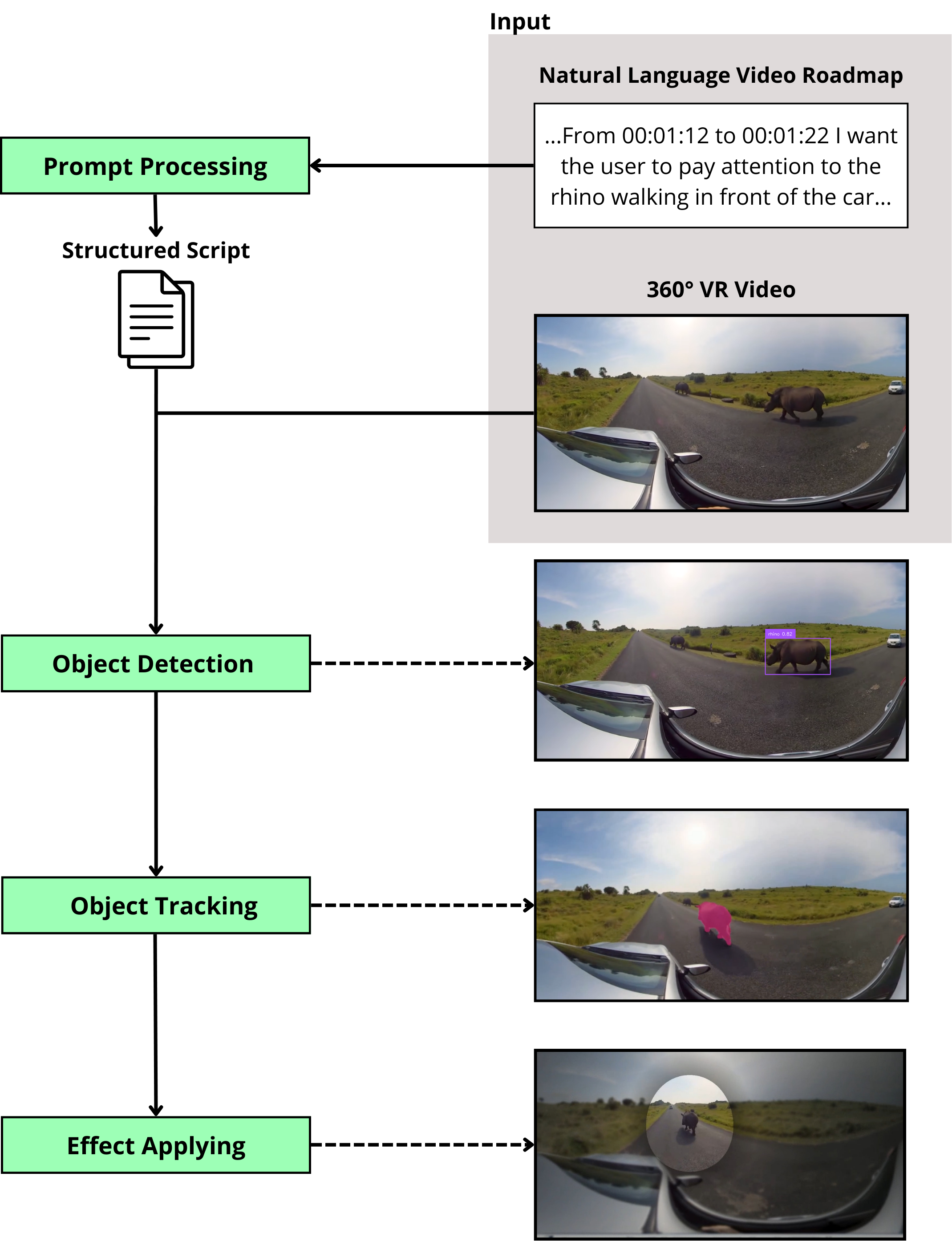}
  \caption{Main pipeline for automatically guiding users' attention during a 360º VR Video.}
  \label{fig:pipeline}
\end{figure}

The proposed architecture, as illustrated in \cref{fig:pipeline}, processes a video according to a user provided road map of what elements the user must pay attention to in each moment of the provided video. First, we process the road map provided by the user, transforming it into a structured script that is composed of the time intervals and element descriptions. Then we iterate over the structured script detecting the target elements of the start of each time interval, segment it and propagate the mask until the end of the interval. Subsequently, the concerned frames are processed by the Effect Applying module, that will carry about the attention attracting.

\textbf{Prompt Processing} module aims to comprehend the user provided road map and transform it into a structured data. The input prompt describes what elements the user must pay attention to each time interval, written in a natural language. This information is propagated to Llama 3 Model \cite{dubey2024llama} which extracts the relevant information and returns a CSV file containing the time intervals and object descriptions.

\textbf{Object Detection} module aims to detect the object described by the structured script on the start of each time interval. The object description and the concerned frame are propagated to Grounding DINO \cite{liu2025grounding}, which detects the object and returns its bounding box that is used as initial tracking.

\textbf{Object tracking} module is responsible for keeping the detected object tracked during the time interval. The bounding box of the object detected at the start of the interval is propagated to Segment Anything 2 \cite{ravi2024sam}, which segments the concerned object and generates a mask for it. Due to its capabilities in video processing, the model can track the target object and propagate its mask over the remaining frames of the time interval.

\begin{figure}[h]
  \centering
  \includegraphics[width=.9\columnwidth]{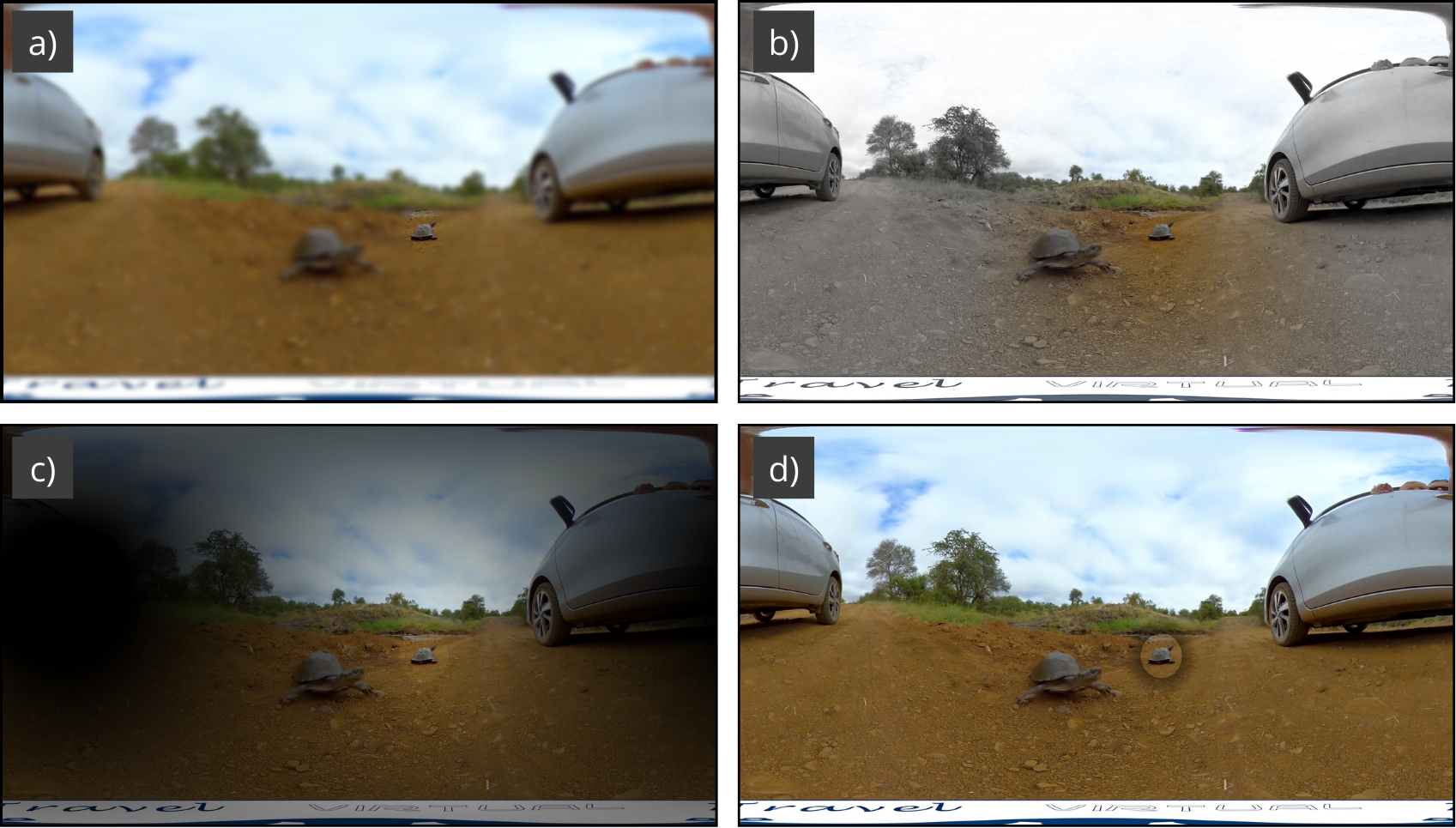}
  \caption{Individual visual effects employed on the combination to direct the users' attention to the farthest turtle. a) Blur. b) Fade to Gray. c) Radial Darkening. d) Halo Darkening. }
  \label{fig:effects}
\end{figure}

\begin{figure}[h]
  \centering
  \includegraphics[width=.9\columnwidth]{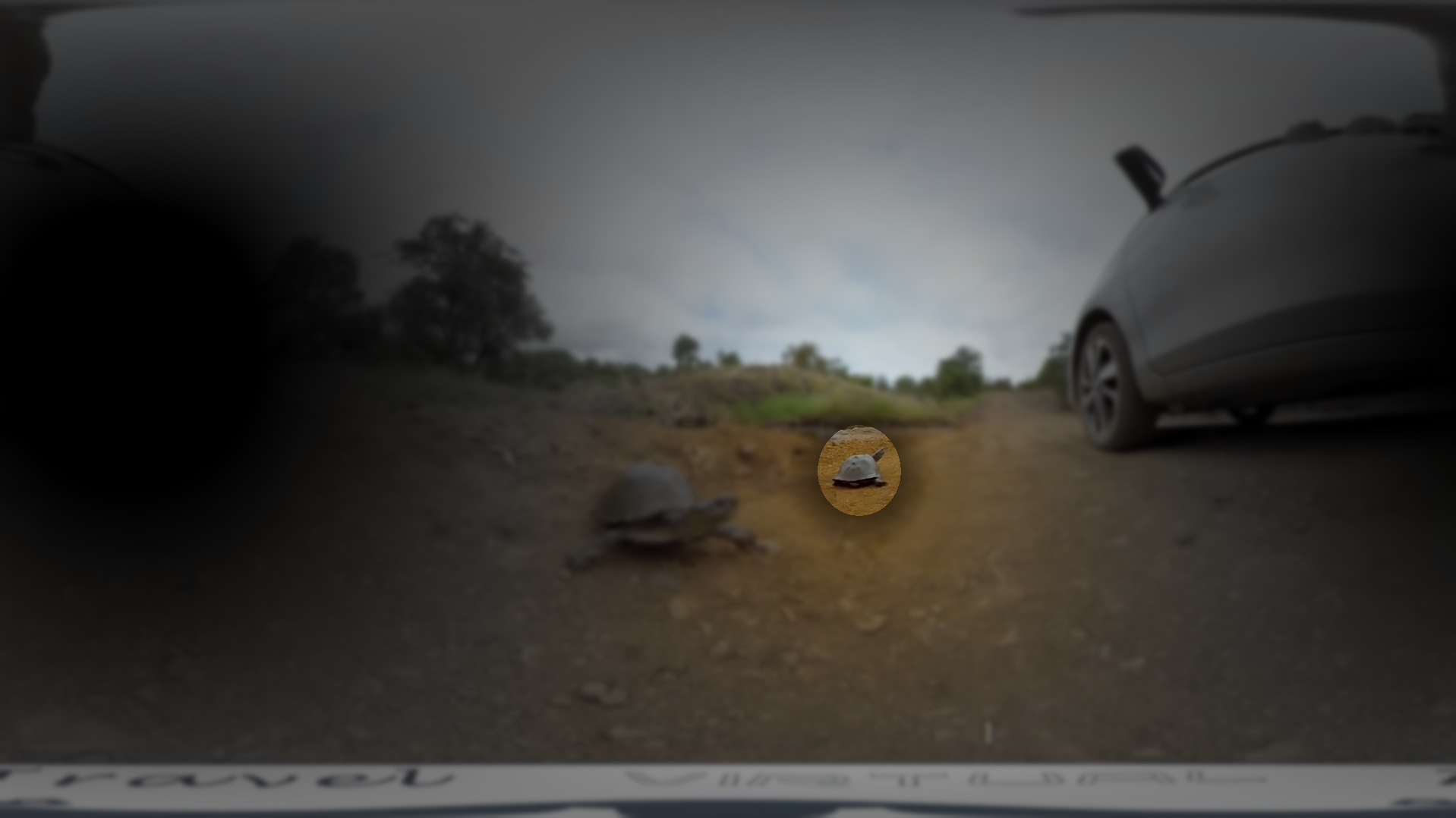}
  \caption{The combination of the four visual effects to direct the users' attention to the farthest turtle.}
  \label{fig:result}
\end{figure}

\textbf{Effect Applying} module aims to, through the application of a combination of visual effects on the frames of the interval, attract the users' attention to the tracked object. Four types of visual effect are combined: Blur, Fade to Gray, Radial darkening and Halo Darkening. The \textbf{Blur} effect is applied to the entire frame, except the radial region around the element of interest, this effect aims to highlight the object of interest in relation to the background and other items present nearby in the scene, doing so through the difference in sharpness, as shown in \cref{fig:effects}-a. The \textbf{Fade to Gray} effect has radial behavior, so that the further away from the central point of the object, the lower the saturation of the pixels present in the image. The main objective is to create a highlight through contrast between the object of interest and the rest of the image, gradually drawing the viewer's attention as the loss of saturation is perceived, as shown in \cref{fig:effects}-b. The \textbf{Radial Darkening} acts similarly to the Fade to Gray, but applies a darkening to the image that increases as you move away from the object of interest. The purpose of this effect is to more objectively guide the user's attention to the region close to the object in cases where they may be looking at the complete opposite or regions very far from the desired element, as shown in \cref{fig:effects}-c. The \textbf{Halo Darkening} effect involves the darkening around the region of interest. It acts inversely to radial darkening, so that the closer to the center of the object of interest, the darker the pixels will be. Because of the unaffected region, the result becomes a thin layer of darkened pixels around the halo formed by the region, helping to distinguish between the main focus and the regions closest to it, as shown in \cref{fig:effects}-d. The combination of the four visual effects is shown in \cref{fig:result}.


\section{Demonstration}
The demonstration is based on a 360º VR video of a Safari Tour in Kruger National Park, in South Africa. The video consists of showcasing some animals while riding a car in the jungle. The 360º camera is located at the top of the car, capturing the entire environment around it. The video processing is realized on an Nvidia RTX 4090 and displayed on a Meta Quest 3.

\section{Conclusion and Future Work}
In this work, we presented Focus360, a novel solution designed to direct users' attention while viewing 360º VR videos. The proposed approach leverages natural language descriptions of the video roadmap to automatically interpret the text, identify key elements, and determine the specific moments that require user focus. By employing a combination of four distinct visual effects, the system effectively guides the user's attention to these elements of interest throughout the video. This integration of visual effects overcomes the limitations of previous methods. In this context, future works will be focused on the evaluation of the proposed method through interviews about the satisfaction and effectiveness in directing users' attention. In addition to comparing the proposed method with other works.

\acknowledgments{
This work has been fully/partially funded by the project Research and Development of Algorithms for Construction of Digital Human Technological Components supported by Advanced Knowledge Center in Immersive Technologies (AKCIT), with financial resources from the PPI IoT/Manufatura 4.0 / PPI HardwareBR of the MCTI grant number 057/2023, signed with EMBRAPII.
}

\bibliographystyle{abbrv-doi}

\bibliography{template}
\end{document}